\documentclass[12pt,a4paper]{article}
\usepackage{graphicx}
\usepackage{caption}
\usepackage{subcaption}

\begin{document}

\title{TiO$_2$-based Memristors and ReRAM: Materials, Mechanisms and Models (a Review)}


\author{Ella Gale$^{1,2}$\footnote{Current Address: School of Experimental Psychology, University of Bristol, 12a Priory Road, Bristol, BS8 1TU, UK. E-mail: ella.gale@bristol.ac.uk}\\
1. Unconventional Computing Group,\\ University of the West of England, Frenchay Campus,\\
Bristol, BS16 1QY\\
2 Bristol Robotics Laboratory, Bristol, BS16 1QY \\
}

\maketitle

\begin{abstract}
The memristor is the fundamental non-linear circuit element, with uses in computing and computer memory. ReRAM (Resistive Random Access Memory) is a resistive switching memory proposed as a non-volatile memory. In this review we shall summarise the state of the art for these closely-related fields, concentrating on titanium dioxide, the well-utilised and archetypal material for both. We shall cover material properties, switching mechanisms and models to demonstrate what ReRAM and memristor scientists can learn from each other and examine the outlook for these technologies.
\end{abstract}

\section{Introduction}

The semiconductor industry has, for many years, managed to increase the number of components per chip according to Moore's law (geometrical scaling). Further advances have been achieved through functional diversification (`More-than-Moore'), namely complementary technologies that increase the usefulness of electronic systems by adding extra functionality. According to the International Technology Roadmap for Semiconductors, to continue at this rate the semiconductor industry needs to combine further miniaturisation with this type of functional diversification, as well as investigate a future beyond CMOS~\cite{322}. There are several technologies that might fit such a desire, two of which are memristors and Resistive Random Access Memory (ReRAM, also known as RRAM). 

In this short review, we shall cover the material properties, elucidated mechanisms and current models in order to explain the state of these fast-moving fields. ReRAM is usually based on transition metal oxides such as TiO$_2$, SrTiO$_3$~\cite{169,170}, NiO~\cite{164}, CuO~\cite{162}, ZnO~\cite{160}, MnO$_x$~\cite{176}, HfO$_{x}$~\cite{352}, Ta$_2$O$_5$~\cite{357}, Ti$_2$O$_{5-x}$/TiO$_{y}$~\cite{353}, TaO$_x$/TiO$_{2-x}$~\cite{359}; both binary and perovskite oxides are capable of resistance switching. Memristors can be made out of TiO$_2$, chalcogenides~\cite{N-777}, polymers~\cite{262,cheapgraphene}, atomic switches~\cite{151}, spintronic systems~\cite{145} and quantum systems~\cite{116}. Biological material and mechanisms like sweat ducts~\cite{276}, leaves~\cite{150,358}, blood~\cite{147}, slime mould~\cite{3,277}, synapses~\cite{41,239} and neurons~\cite{247} can be described as memristive. In this review we shall restrict our focus to TiO$_2$-based memristors and ReRAM because TiO$_2$ is the material 
that has received most work in memristors and is a well-studied and archetypal system 
within ReRAM~\cite{136}. We shall start by going through a brief the history of the fields, then the materials and mechanisms before looking at the models in more detail.

\begin{figure}[!tbp]
    \centering  
    \includegraphics{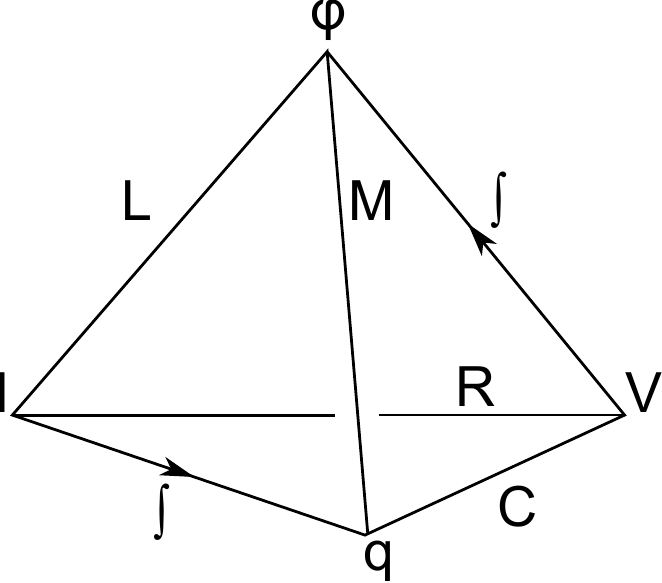}
    \caption{The four circuit measurables and the six relationships between them.}
    \label{fig:tetra}
\end{figure}

The resistor, capacitor, and inductor are the three well-known fundamental circuit elements, discovered in 1745, 1827 and 1831 respectively. Based on an assumption of completeness (see figure~\ref{fig:tetra}), in 1971 Leon Chua postulated a 4$^{\mathrm{th}}$ fundamental circuit element~\cite{14} which would relate charge $q$ to magnetic flux $\varphi$ and would have the distinction of being the first non-linear circuit element (where the non-linearity arises because $q$ and $\varphi$ are integrals of the circuit measurables, current, $I$ and voltage, $V$). No physical instantiations of the memristor were generally acknowledged until 2008 when Strukov et al. realised that their molecular electronic switches' behaviour was due to the titanium electrodes and not the organic layer~\cite{330} and announced that they had found the memristor~\cite{15}. At that time, it was believed that memristors could not have been made contemporary with the other fundamental circuit elements as memristance was a nanoscale 
phenomenon~\cite{15}, `essentially unobservable at the millimetre scale' and above~\cite{330} (since experimentally disproved by the fabrication of macroscopic memristors~\cite{macro1,macro2,macro3,M0}). It later became apparent that memristor-like devices had been fabricated before 2008: resistance switching was first observed in oxides in a gold-silicon monooxide-gold sandwich in 1963~\cite{350}, the first metal oxide resistance switch was reported in nickel oxide in 1964~\cite{351}, memristor-like switching curves were observed in TiO$_2$ thin films in 1968~\cite{183} and the memistor~\cite{29}, a 3-terminal memristive system, was fabricated in 1960 (although is disputed as being an example of a memristive system~\cite{312}). These examples inspired some searches for the earliest known memristor~\cite{119,236} with the singing arc~\cite{316} in 1880 and the coherer in 1899~\cite{313} being the oldest devices identified so far.

When the memristor was announced the field of ReRAM was already mature: two contemporary reviews~\cite{296,155} summarise the state of the art. ReRAM is generally made of a metal-insulator-metal structure, typically using transition metal oxides as the insulator, while the electrodes tend to be a noble metal, with Pt being a popular choice (but Al, Au, ITO (Indium Tin Oxide) and Si are all also common). ReRAM has two resistance states: a high resistance state (HRS) and low resistance state (LRS). It also has two switching modes that go between them. The simplest to understand is bipolar resistance switching (BRS) where the device switches depending on the magnitude and direction of the applied voltage. The other is unipolar resistance switching (UPS) which depends only on the magnitude of the voltage. Both BPS and UPS $V$-$I$ curves resemble memristor curves, a fact which has led to much debate in scientific literature. In this review, we shall compare and contrast the approaches of both fields, using 
titanium dioxide devices as examples. 

\section{Materials} 

One reason for the confusion between the fields of memristors and ReRAM is the different standard measurements. The term ReRAM refers to a suggested use for a range of materials (namely as memory), whereas memristors are named after a property and have been investigated not just for RAM, but also for such uses as neuromorphic hardware (e.g.~\cite{71,239,243,N-442,314,116,DavidJ1}), chaotic circuits (e.g.~\cite{40,252,253,251}), processors (e.g. ~\cite{N-197,UCNC,processor1,processor2,processor3,maze}), vision processors (e.g.~\cite{239,275,TAROS,323,image1}) and robot control (e.g.~\cite{A2,R1,DavidJ1}) to name a few. Memristor measurements tend to be a.c. measurements, often done at different frequencies. This is partly because initially the memristor was thought to be a purely a.c. device and only recently have the d.c. aspects of the device received attention~\cite{N-442,N-647,SpcJ,242}. ReRAM measurements are reported in different ways depending on the type of device. BPS devices are measured in a 
similar way to memristors; but UPS devices are taken to a high voltage with a compliance current set, which is the method of forming the device. After forming, the device is then taken around a positive voltage loop from low to high voltage (where the high voltage is usually below the forming voltage). As UPS ReRAM is not subjected to bipolar voltage waveforms, it is difficult to compare them to memristors.


The first device called a memristor was made from TiO$_2$ thin film~\cite{15}, but the first TiO$_2$ thin film device that demonstrated memristor curves was fabricated in 1963~\cite{183}, this work showed the film pass through a family of hysteretic curves during repeated tests: from pinched open curves (resembling a jelly-bean in shape), to BPS curves before collapsing to a single line as the device degraded. Resistance switching in TiO$_2$ materials had been seen several times in ReRAM experiments before 2008, for example: sputter-grown TiO$_2$ with Pt electrodes~\cite{157} and Pt-TiO$_2$-Ru stacks~\cite{159}. But it was not until Hewlett-Packard's announcement that people became interested in making memristors, and so the report in 2009~\cite{28} of a flexible Al-TiO$_2$-Al memristor fabricated using solution processing garnered a lot of interest (a flexible ReRAM equivalent came out the same year~\cite{194}). The use of aluminium electrodes however, did suggest that aluminium oxide might be involved in 
the 
switching: XPS and EELS has shown the existence of a layer of Al$_2$O$_3$ in Al-TiO$_2$-Al ReRAM BPS devices~\cite{174}, Al$_2$O$_3$ had been added to organic memory devices to improve switching~\cite{229} and been implicated in resistance switching~\cite{178,175,177,189}. There is evidence for Al being supportive in TiO$_2$ resistive switches grown by atomic deposition~\cite{159} and there is even intriguing evidence for a mixed phase~\cite{318}. In~\cite{28} the authors state that the devices still switch if the electrode materials are changed to a noble metal (Au) and~\cite{260} showed that changing the device electrode changes the form and reversibility of the switching. Other work has concentrated on making cheaper and easier to manufacture memristor devices ~\cite{M0,103,cheap2}, but presently the memristor is a difficult device to manufacture for the purposes of experimental testing, which has led to the field being centred around simulation and theoretical work.

\section{Mechanisms}

Many suggested mechanisms have been put forward to explain the causes of resistance switching and of memristance, and TiO$_2$-based devices have examples of all of them. The high resistance material is generally believed to be stoichiometric TiO$_2$~\cite{15}, but there is debate about what the low resistance material is, and, given that the memristors/ReRAM are made in several different ways there is no expectation that there is a single explanation or material cause.

The most popular suggested mechanisms can be split into two groups: ionic and thermal. 

\subsection{Ionic}

The ionic mechanism in TiO$_2$ devices involve the migration of oxygen vacancies (although suggestions of migrating OH$^{-}$ ions have been made~\cite{175,94}). This movement of vacancies creates auto-doped phases which are metallically conducting for TiO$_{(2-x)}$ for $x>1.5$~\cite{332}. The oxygen ions may combine at the anode and evolve O$_2$ gas, this has been seen in ReRAM~\cite{158}, the Strukov memristor when fabricated with macroscale electrodes~\cite{225} and the sol-gel memristor~\cite{260}. There is some evidence that the choice of electrode material can hinder the production of O$_2$, for example, it is thought aluminium supports TiO$_2$ memristance by acting as a source or sink of oxygen ions~\cite{174,130}. Most memristor-based modelling has concentrated on modelling the flow of these oxygen vacancies (see later). \cite{297} suggests that switching is dominated by ionic motion rather than charge trapping. A mixed mechanism involving oxygen vacancy transport between the tip of a conducting 
filament (itself formed by thermal mechanism) and an electrode has been suggested~\cite{329}. Alternative electrochemical mechanisms that explain memristance via a change in titanium oxidation state, Ti(IV)$\rightarrow$Ti(III), have been put forward~\cite{94} although not widely adopted. 

\subsection{Thermal}

Joule heating is the process where the application of an electric field and flowing current heats the material and changes its structure. TiO$_2$ atomic deposition thin films can form conducting filaments as extended defects along grain boundaries~\cite{159-15}, the  ions drift, forming a path which breaks with excess heat (i.e. too high a voltage) and can be reformed via the same mechanism. This mechanism has been credited with causing both HRS and LRS states~\cite{159}. Single crystals of SrTiO$_3$ show switching in the skin after being subjected to an electric field under ultra-high vacuum. This causes `dislocations' which are broken by ambient oxygen and then reformed via electroforming. This is present in a crystalline structure and suggests a poly-filamentary mechanism~\cite{158}. The conduction channels are usually ohmic and it has been shown that the resistance and reset current bears little relation to the composition of the material or whether it is operated in a unipolar or bipolar mode, which 
suggests, and has been verified by simulation, that the mechanism is thermally activated dissolution of conducting filaments~\cite{168}. This could explain why compliance current choice controls the resistance and reset current and suggests that larger resistance values are more related with the heat sink effects of the metal oxide, rather than its electronic properties.

The form of these conducting filaments / channels is debated. In 2009 A Pt-TiO$_2$-Pt ReRAM showed a Magn\'{e}li phase~\cite{130}, Ti$_n$O$_{2n-1}$, crystallographic shear plane from Rutile~\cite{317}, roughly 10-20nm in diameter and which shows metallic (i.e. ohmic) conduction. Magn\'{e}li phases are considered likely to be the material cause as they are more thermodynamically stable than vacancies spread through-out the material~\cite{298}. Single crystal TiO$_2$ nanorods have shown bipolar switching~\cite{166} which have a rectifying effect. Conical Magn\'{e}li phase filaments have been observed and shown to be the cause of fusing and anti-fusing BPS~\cite{165}. Hourglass-shaped Magn\'{e}li phases were seen in~\cite{130}, which were formed by alternating polarity voltage and proved to be more stable in operation. Fractal conducting filaments, have been suggested by simulation~\cite{186}, observed~\cite{333} and are thought to relate to dielectric breakdown~\cite{219}. From spectromicroscopy and TEM, 
electroforming the Strukov memristor has been found to produce an ordered Ti$_4$O$_7$ Magn\'{e}li phase~\cite{226}.

\subsection{Other mechanisms}

Other mechanisms have been put forward. The UPS switching has credited to: metal-semiconductor transitions~\cite{331}, crystalline TiO$_2$-amorphous TiO$_2$ phase transition via conduction heating and breaking~\cite{159}, raising and lowering Schottky barriers via bulk (UPS) or interface (BPS) transport of oxygen~\cite{169}, and conductance heating causing lateral transport of conducting filaments~\cite{168}. There is also a debate as to what the structure of the less-conducting thin-film TiO$_2$ is, with rutile ($r$-TiO$_2$)~\cite{159} and amorphous titanium dixoide ($a$-TiO$_2$) being the most popular suggestions. Note, that $a$-TiO$_2$ has also been suggested as the conducting form of TiO$_2$ and that Magn\'{e}li phases are sheer planes from $r$-TiO$_2$.

Currently, the Magn\'{e}li conducting filaments are the most generally accepted mechanism, but as all the materials are treated differently, measured differently, and titanium dioxide has a wealth of behaviours, it is unlikely that there will emerge only one explanation for all published TiO$_2$ devices.

\section{Models}

Generally, ReRAM modelling has concentrated on models based on the materials science and chemistry of the devices (and are described above), whereas the memristor, by virtue of being predicted from electronic engineering theory tends to include more physics- and mathematics-based approaches. 

\subsection{Models based on materials science}

Chua's 1971 paper~\cite{14} made the observation that there were only five relations between the four circuit measurables: charge, $q$, magnetic flux, $\varphi$, current, $I$ and voltage, $V$. Two relations were the definitions of charge and flux (as $I = dq/dt$ and $V=d\varphi / dt$), the other three were the constitutive definitions of the resistor ($V=IR$), capacitor ($q=CV$) and inductor ($\varphi=LI$). From these facts, the missing 6$^{\mathrm{th}}$ relationship would be $\varphi = M q$, see figure~\ref{fig:tetra}. When Strukov et al. announced their discovery of a memristor~\cite{15} they included a simple model based on treating the memristor as two space-conserving variable resistors in which they considered the velocity of the boundary $w$ between TiO$_2$ and TiO$_{(2-x)}$ phases. This model has been both highly criticised and widely used, both as is and with extensions. 

The model as reported~\cite{15} did not include the magnetic flux which is expected from Chua's definition~\cite{14}. This lead to questions, including the question of whether a memristor had actually been fabricated~\cite{142}. Attempted solutions to this issue have included fixes to the model~\cite{F0c}, statements that the magnetic flux was a theoretical construct~\cite{15,334} and not related to a material property~\cite{119} or that it was the non-linear relation between $I$ and $V$ that defined the device~\cite{15}. The Strukov model as presented assumes linear dopant drift under a uniform field~\cite{15}, which is widely believed to be unlikely in devices with high electronic field~\cite{86}. Several people have undertaken to fix this:~\cite{2} introduced the concept of a rectangular `window' function that would prevent $w$ from going out of its bounds (0 and $D$ -- the thickness of the semiconductor layer) and slowed the rate of motion down near the boundary edges (making it non-linear), \cite{46} 
improved on this to prevent $w$ from getting stuck at the boundaries and~\cite{306} adjusted it to allow the maximum to be tuned to less than 1, thereby introducing more variability in device modelling, \cite{328} concentrated on applying the models to nanostructures and ~\cite{327} concentrated on neuromorphic systems. A further confusion arises from the widely-interpreted implication that the uniform field is across the entire device suggested by the wording in~\cite{15}, which is not what was intended~\cite{St1}. A field with a discontinuity is problematic from the electrochemical point of view~\cite{294}, especially as the discontinuity is located on $w$~\cite{289}, and may require further work with windowing functions.

Despite these issues, the Strukov model has been highly adopted, and most work involves simulating with it or improving it with window functions. An in-depth discussion of the uses of it is beyond the scope of this review, but it has generally been used to model test circuits, often using SPICE (Simulation Program for Integrated Circuits Emphasis, a standard electronic engineering simulation package) and comparisons of different SPICE implementations of the models described in this section are given in~\cite{345,346}. Three models have extended it, ~\cite{83} demonstrated a SPICE model with the initial state as a variable, Georgiou et al. extended the model using Bernoulli formulation, introducing extra parameter to get a measure of the device hysteresis~\cite{224} and ~\cite{348} extended it by demonstrating a SPICE model of a magnetic flux-controlled memristor.

Other models have been proposed for use in modelling TiO$_2$ devices based on more realistic models of the materials. A model based on the idea of setting $w$ to the tunnelling barrier length between TiO$_{(2-x)}$ and the electrode was derived by experimentally fitting to the Strukov memristor after electroforming had created a localised conducting channel~\cite{220}. This model gives good agreement with experimental data~\cite{220}, but has the drawbacks of being difficult to simulate (containing a hyperbolic sine term and two exponential terms) and uses 8 fitting parameters for which the experimental analogues are not completely clear (although they are related to nonlinearities due to the high field and Joule heating~\cite{305}). The data in~\cite{220} has been used to create other models: \cite{307} presented a version implemented in SPICE; \cite{305} presented a more general model with a threshold that was easier to simulate; \cite{308} presented a tunnelling model that required only one fitting 
parameter, which has also been implemented in SPICE~\cite{320}; \cite{310} presented a tunnelling model that included a threshold relation and was applied to both memristors and ReRAM. A good model requires that `any parameters in the model determined by fitting to experimental data should be intrinsic to the device'~\cite{335}. There is a trade-off between the quality of the fit, which can require many fitting coefficients; and the relation to physical processes, which requires that each fitting coefficient is strongly related to a real-world property. 

The trend in building memristor device models is currently towards the more general (in that they can model more devices and types of devices), experimentally-informed (in that the model can be understood in terms of the physical processes happening in the device), easy-to-simulate models.    
                                                                                                                         
\subsection{Models based on electronic engineering}

As Strukov's model is phenomenological, so the derivation of Chua's model was more rigorous and mathematical, nonetheless, the theoretical idea of a memristor has not escaped the influence of real devices. Before a memristor was identified, the concept of a memristor was expanded to that of a memristive system~\cite{84}, which has two state variables, this concept was used to describe a thermistor and neuron ion channels. Since the discovery of a memristor device, the terminology has changed to refer to memristive systems as a type of memristor, rather than vice versa. In 2011, Chua published a paper~\cite{119} in which he asserted that all ReRAM devices with pinched hysteresis loops are memristors. In~\cite{119} Chua suggested that experimentalists should record $q$-$\varphi$ data because $V$-$I$ curves are open to effects of input voltage (a point well illustrated in~\cite{224}, which offers an attempted theoretical solution) and are not predictive. The focus shifted in~\cite{279} where the frequency 
dependence (which is mentioned in the 1971 paper~\cite{14}) is elevated to being two of the three identifying `fingerprints' for experimentalists to search for (namely that the hysteresis loop shrinks as the frequency tends to infinity). Further expansions of the memristor idea have been undertaken. The memristor is a passive device, but the idea of an active memristor has been shown to be useful~\cite{251,277,40}. The pinched hysteresis loop, the most distinctive property of the memristor, has been relaxed to a non-self-crossing pinched hysteresis loop~\cite{291,93,276} (matching experimental data for memristors~\cite{183,255} and for ReRAM~\cite{292}). These changes have led to some criticism~\cite{321} that theoretical ideas cannot be redefined at whim to fit experimental data. This is an interesting point to ponder. A theory that has no relevance to real world systems is of little practical use, and adjusting the presentation and switching which properties are necessary to define device behaviour and 
which are merely 
indicative in response to a larger group of discovered devices seems, to me at least, to be entirely appropriate for theoretical science. Nonetheless, none the changes in the definition of the memristor put forward by Chua have changed the basic concept, that of a non-linear resistor which relates the time integral of the current to the time integral of the voltage and produces pinched (although now not necessarily self-crossing) $V$-$I$ curves which shrink with increasing frequency.

\section{Memristors and ReRAM - one field or two?}

Having presented an overview of both fields it is worth asking if they are one field or two. Initially the two communities ignored each other and thus several memristor papers published early on presented phenomena already discovered within ReRAM (as described earlier) and, even now, some ReRAM papers fail to reference memristor work~\cite{338}. Generally, opinion has shifted to ReRAM scientists using and extending~\cite{292} memristor theory, while memristor scientists look to ReRAM to explain device properties~\cite{240,309}, suggesting that even if they are not the same field there is at least convergence between them. Some researchers are even using the terms interchangeably~\cite{339} and some are taking memristors as just another type of technology with which to build ReRAM~\cite{364}.

\begin{figure}[!tbp]
    \centering
    \begin{subfigure}[]{0.4\textwidth}
        \includegraphics[width=\textwidth]{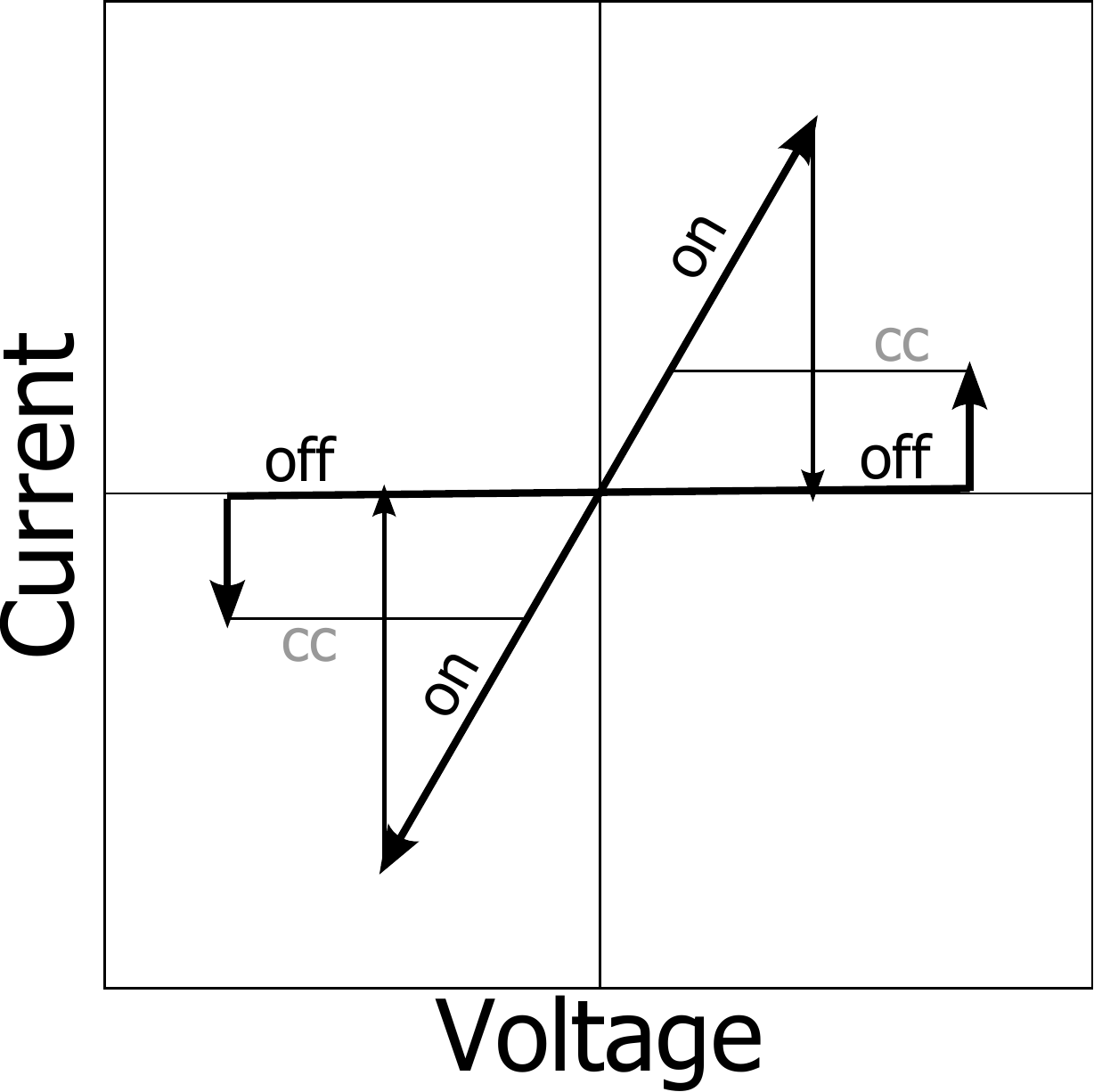}
        \caption{Typical unipolar ReRAM}
        \label{fig:unipolar}
    \end{subfigure}
    \begin{subfigure}[]{0.4\textwidth}
        \includegraphics[width=\textwidth]{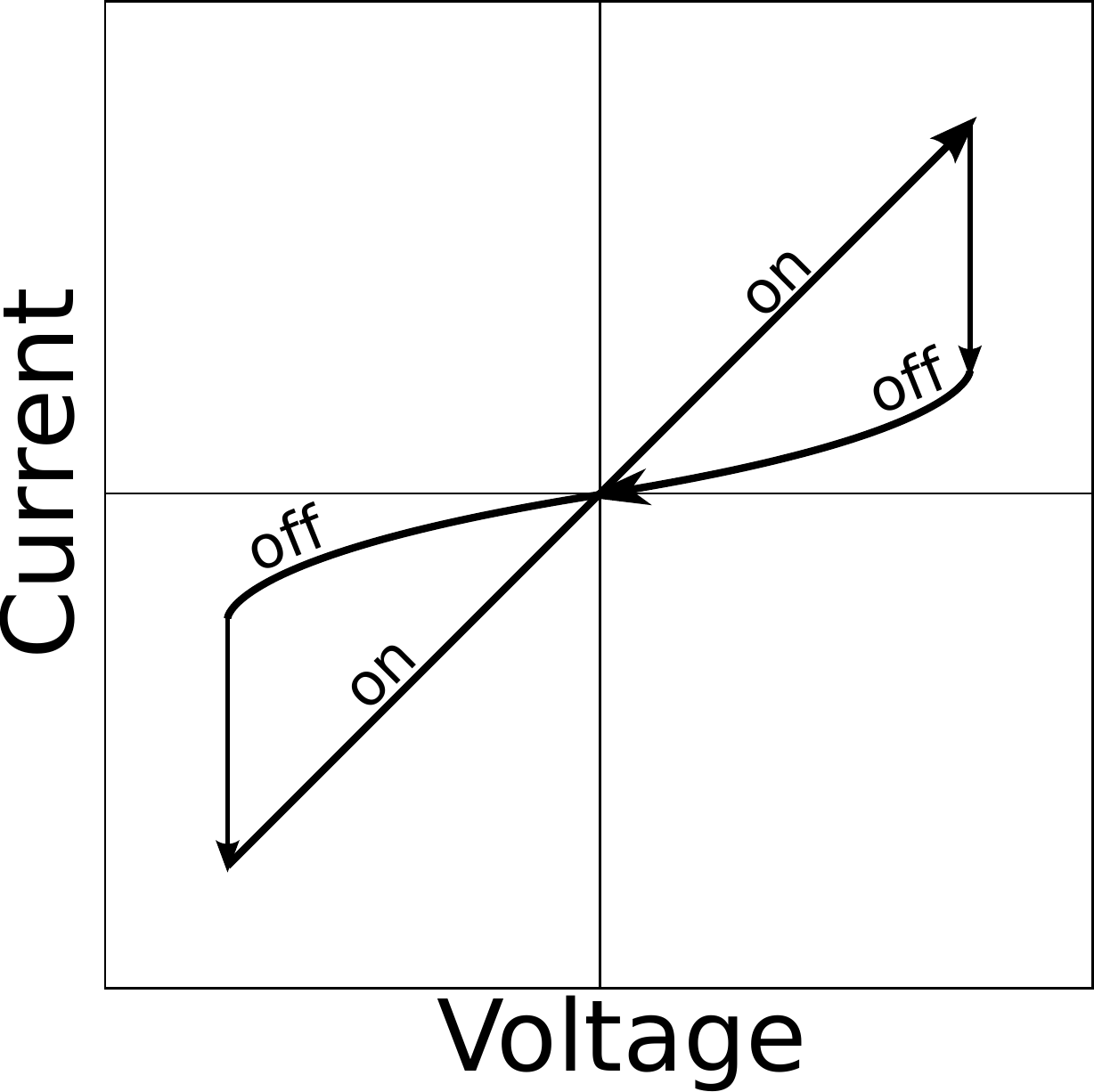}
        \caption{Typical bipolar ReRAM}
        \label{fig:bipolar}
    \end{subfigure}
    
    \begin{subfigure}[]{0.4\textwidth}
        \includegraphics[width=\textwidth]{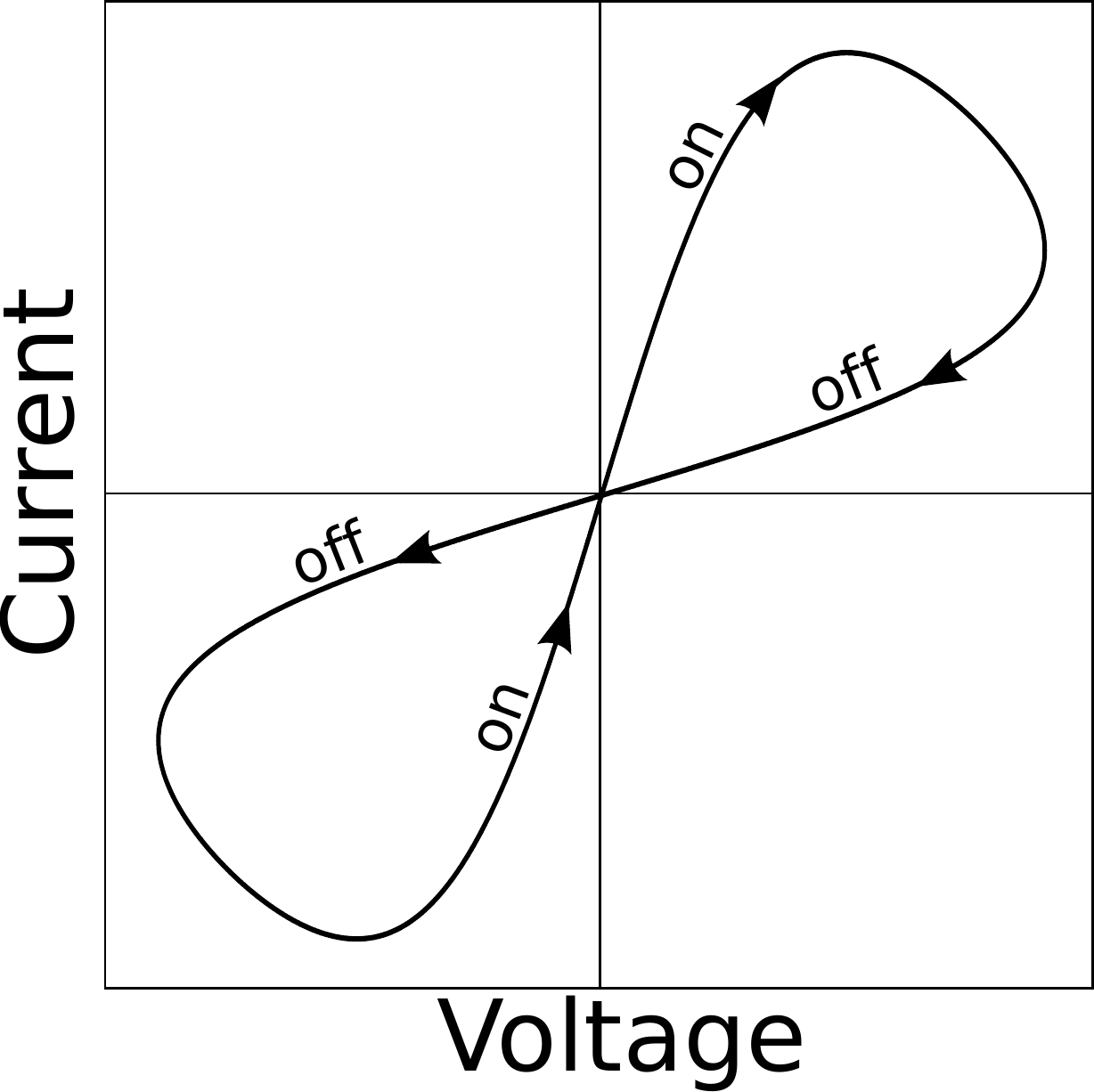}
        \caption{Theoretical Memristor V-I}
        \label{fig:theorymem}
    \end{subfigure}
    \begin{subfigure}[]{0.4\textwidth}
        \includegraphics[width=\textwidth]{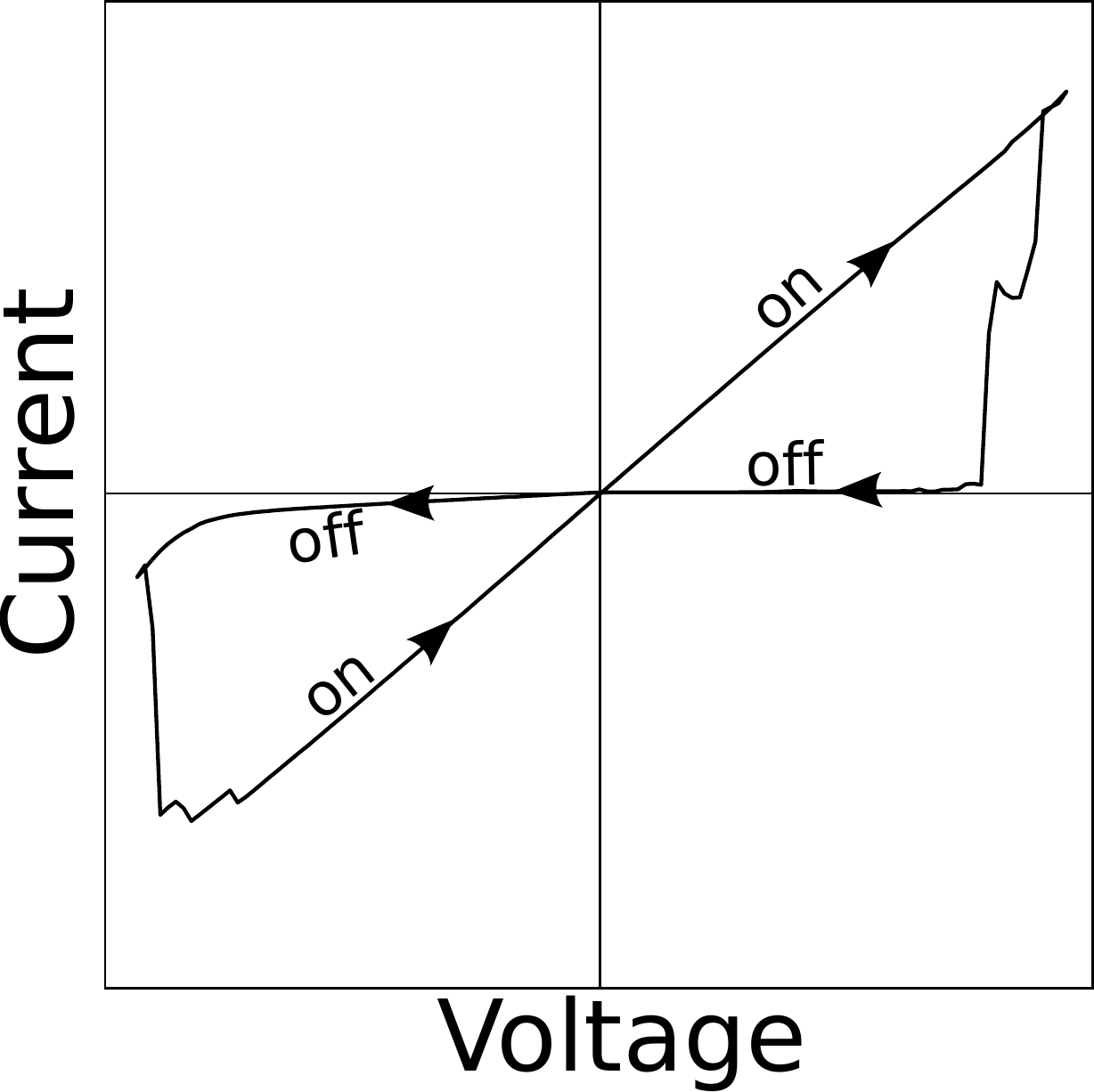}
        \caption{An Example Memristor curve~\cite{260}}
        \label{fig:realmem}
    \end{subfigure}
    \caption{Predicted and typical ReRAM and Memristor curves.Note that the example from~\cite{260} switches from on to off, many memristors switch the opposite way round, and would go around the I-V curve in the opposite direction}
\end{figure}

Questions highlighting the differences remain however: The memristor is defined as a nonlinear, analogue device; so devices that undergo ohmic conduction (as seen in conductive filaments) are not strictly memristors, although it is possible to model them by including conducting channels in the device model~\cite{254}. However, the memristor equations at certain frequencies do give more triangular shaped curves~\cite{15} which can match the observed shape (
Figure~\ref{fig:realmem}.). To match memristor theory (Figure~\ref{fig:theorymem}) to real devices, different extensions to the theory, such as non-zero crossing~\cite{292}, conducting filaments~\cite{254}, active memristors~\cite{251} and analysis of use cases such as reading and writing operations of memory~\cite{339,340} are required. However, memristor theory is so elegant and useful (for example, it neatly explains the behaviour of $I$-$V$ loops shrinking to a single valued function that has been observed in ReRAM~\cite{183,159}), it seems worth the effort to add real-world considerations.

The question of whether the work belongs in one or two fields is far from a settled one. It seems, however, that BPS ReRAM (Figure~\ref{fig:bipolar}) is a real-world instantiation of the memristor and whether UPS ReRAM (Figure~\ref{fig:unipolar}) is considered a memristor depends which formulation of the memristor definition you adopt: whilst in the ohmic conducting regime UPS ReRAM does not satisfy the 1971 definition~\cite{14} of the memristor, because the memristance is not changing with $q$. The system can, however, be described as 
a memristive system as per the 1976 definition~\cite{84} with the state of conducting filaments (i.e. fused or connecting) as a second state variable. Under the nomenclature put forward in~\cite{119} these devices can thus be considered memristors.

\section{The Future}

The memristor has only been understood as an existent device for five and half years and in that time it has drastically effected the outlook for ReRAM devices by providing a deeper theoretical understanding of aspects of their operation and suggesting uses for these devices other than merely as a novel type of storage. In turn, the rich history and vast amount of work on the materials, their properties and how they can be controlled that comes from ReRAM scientists have allowed the memristor concept to move from interesting phenomenon to practical device far quicker than would otherwise have happened. I suspect that the trend for ReRAM and memristors to be increasingly be viewed as describing the same phenomena will continue.

Currently, work is going in several directions. The first computer containing ReRAM memory (based on TaO$_x$) was  announced by Panasonic in 2013, but many people and a few companies, are racing to produce their competing commercial memristor/ReRAM memory, be it computer RAM, flash drives or other form of storage, and so there is a lot of work on stabilising and controlling the device's properties and looking for other materials for manufacturing. 

The first commercial suggestion for memristors was memory because the memristor's small feature size could offer an increase in memory density, wouldn't require power to hold a value (so it is low energy electronics) and could be used for multi-state memory. Alongside the usual manufacturing faults, memristors suffer some unique issues on which there has been much work and many attempted fixes, both based on memristor theory~\cite{340,339} and experiment~\cite{341}. For example, according to the theory, memristor state will change when read, affecting data integrity, this can be solved by using a read pulse below a threashold time~\cite{340} or a reading algorithm that reads and then rewrites the memory cells~\cite{362} (this is also useful for reducing sneak-path-related read errors). Other work has focused on improving memory tests to identify undefined states~\cite{339} and fabrication errors ~\cite{344} (by making use of sneak paths to test several memristors at the same time). A big problem for 
memristor-based cross-bar memory is sneak paths which can result in erroneous memory values being read and are a big problem for shrinking memory size. Attempted solutions include: Hewlett-Packard's reading algorithm~\cite{362} described above; unfolding the memory (only one memristor per column/row, which greatly reduces the possible gains from shrinking memory size; adding an active element like a diode~\cite{356}, which might add delay, or a transistor gate~\cite{364}, which would remove the advantage of the memristor's small size; using anti-series memristors as the memory element so that the total resistance is always $R_{\mathrm{on}} + R_{\mathrm{off}}$~\cite{361}, this would require differentiating between $R_{\mathrm{on}}-R_{\mathrm{off}}$ and $R_{\mathrm{on}}-R_{\mathrm{off}}$ for logical values; using multiple access points~\cite{343}; using a.c. sensing~\cite{360} at the cost of increased complexity; making use of the memristors own non-linearity (instead of a diode)~\cite{359,341}; making use of 
a 3-terminal memristor~\cite{342} (instead of a transistor) which forces sneak paths to have a higher than $R_{\mathrm{off}}$ resistance (which is the best case failure for sneak paths) but requires an additional column line. The authors of~\cite{342} concentrated on the memistor~\cite{29} as an example device, but intriguingly the idea could be applied to the 3-terminal plastic memristor~\cite{12,45}. For memristor-based memory to be adopted, there is much further work to be done in this area, especially on experimentally testing these approaches. 

Another big area is the design and fabrication of neuromorphic computers using memristors, which will require novel approaches and hardware instantiations. Because memristors can implement IMPLY logic, Bertrand Russell's logical system~\cite{Russell} is getting something of a renaissance. If memristors natively implement IMPLY logic~\cite{242}, perhaps as spikes~\cite{Spc2,UCNC}, then there will be a lot of work on engineering memristors into current circuit design approaches (which traditionally use AND, OR, NOT, NAND and similar). Several workers have suggested doing digitised stateful logic in CMOS-compatible cross-bar memory arrays (using the IMPLY and FALSE operation set) and started to look at design methodologies: as in~\cite{348}, making encoders/decoders out of transistor-memristor arrays~\cite{349} and making use of the parallelisability of the architecture for simultaneous bitwise vector operations~\cite{354}. Recently, several researchers have started to look at hysteresis~\cite{281,282,224,324} 
and other figures of merit~\cite{325,326,306} of practical use to engineers.

An interesting and unexpected outcome of this work is the discovery that evolution has made use of memristors and memristive mechanisms, where living memristors have been identified in leaves, skin, blood, and eukaryotic mould so far, and memristor theory has been used to understand learning in the synapses and simple organisms and to update neuron models such the Hodgkin-Huxley~\cite{247}. This could suggest that electrophysiology is a field ripe for memristor research and perhaps the most ground-breaking memristor work will come from linking the manufactured devices with living memristors, either directly or via the creation of bio-inspired computers which may usher in an entirely new paradigm of computational approaches.

\section*{Acknowledgements}

This work was supported by EPSRC on grant EP/HO14381/1. The author would like to thank Oliver Matthews and Ben de Lacy Costello for useful discussions.

\section*{References}
\bibliography{UWELit}
\end{document}